\documentstyle[12pt]{article}
\begin{document}
\setlength{\baselineskip}{0.15in}
\newcommand{\beq}{\begin{equation}}
\newcommand{\eeq}{\end{equation}}
\newcommand{\bi}{\bibitem}

{\hbox to\hsize{May, 1996  \hfill TAC-1996-010}

\begin{center}
\vglue .06in
{\Large \bf {Antimatter in Different Baryogenesis Scenarios
 }}\\
{\it{presented at International Workshop on Baryon Instability,\\
Oak Ridge, Tennessee, March 28-39, 1996.\\
(to be published in the Proceedings)}}\\[.5in]
\bigskip
{\bf A.D. Dolgov}
 \\[.05in]
{\it{Teoretisk Astrofysik Center\\
 Juliane Maries Vej 30, DK-2100, Copenhagen, Denmark\\
and\\
ITEP, Bol. Cheremushkinskaya 25, Moscow 113259, Russia.}}\\[.15in]

\end{center}
\begin{abstract}
Possible mechanisms of abundant creation of antimatter in the universe
are reviewed. The necessary conditions for that are: baryonic  charge
nonconservation, spontaneous breaking of charge symmetry or nonequilibrium
initial state, and the formation of appropriate initial conditions during
inflation. In this case the universe may be populated with domains, cells,
or even stellar size objects consisting of antimatter.
\end{abstract}
\section{Introduction}

The problem that I am going to discuss is not directly related to the subject
of this conference dedicated to experimental search of baryon nonconservation.
Still there is one thing in common, all models of cosmological creation of
antimatter request nonconservation of baryonic charge. There may be of course
a production of antibaryons by e.g. decays
or annihilation of new long-lived heavy
particles (like quasistable neutralinos of supersymmetric models) which may
proceed with baryonic charge conservation or even production of antinucleons
by energetic cosmic rays but this is not what is usually
understood as creation of antimatter.

We know from observations that the universe in our neighborhood is 100\%
charge asymmetric. There are only baryons and electrons and no their
antiparticles in a comparable amount. Though the asymmetry is large now, in
some sense it is very small. The number density of baryons, $N_B$,
relative the number density of photons in the cosmic microwave background
radiation, $N_{\gamma}$ is:
\beq{
N_B/N_\gamma \approx 3\times 10^{-10 \pm 0.3},
\label{nbngamma}
}\eeq
This means that the universe was almost charge symmetric at high temperatures,
$T>(a\, few)\times 100$ MeV. At these temperatures the excess of baryonic
charge was approximately one unit per $10^9$ baryons. Still though the ratio
(\ref{nbngamma}) is very small, it is 9 orders of magnitude
larger than it would be in the case of locally charge symmetric universe.
We do not know if all the universe is charge asymmetric with the same universal
magnitude of the charge asymmetry or the charge asymmetry is point dependent
and can even change its sign. Nothing is known about the
size of these locally asymmetric domains, $l_B$. Existing data indicate
that $l_B > 10$ Mpc. Whether $l_B$ is above or below the present day
horizon, $l_h = 10$ Gpc,
is an intriguing question and in what follows I will discuss the
models which predict a relatively small value of $l_B$, so that antimatter may
be accessible to observations.

It is very important for all these models as well as for the
planned experiments on search of baryon nonconservation
to know if baryonic charge is indeed nonconserved.
At the present time cosmology gives the only "experimental" and a very strong
argument in favor of nonconservation of baryons. In other words our existence
strongly implies baryon nonconservation. This is not just that the
baryon asymmetry of the universe can be generated only if baryonic charge is
nonconserved as was suggested 25 years ago by Sakharov \cite{ads}. (For
possible
but rather exotic exceptions see review paper \cite{ad1}.) There is something
more, namely that sufficiently long inflation could not go
with conserved baryonic
charge \cite{dsz,ad1}. Since it seems that without inflation is impossible to
make a suitable for life universe we have to assume that baryons are indeed
nonconserved. The argument goes as follows. For successful solution of
cosmological problems\cite{guth}  inflationary stage should last sufficiently
long (for a review see e.g. books \cite{adl,dsz}). The duration
of inflation $\tau$ should be larger than 60 Hubble times, $H_I\tau >60$,
where $H_I$ is the Hubble constant during inflation such that the scale
factor, which describes the universe expansion, behaves as
$a(t) \sim \exp(H_I t)$. One may say that in order to create the observed
number density of baryons, the initial baryonic  charge density
at the onset of inflation should be unnaturally
large, at least $e^{180} (T_{Rh}/2.7K)^3$ times
larger than at the present day. Here 2.7K is the temperature of the
cosmic microwave background radiation today and $T_{Rh}$ is the temperature
at the end of inflation. Such a large number
is of course not natural but it does not mean impossible. What makes
inflation with conserved baryons impossible is the
energy density considerations.
The Hubble parameter is expressed through the
cosmological energy density density $\rho$ as
$H= \sqrt{ 8\pi \rho / 3m_{Pl}^2}$. To make an exponential expansion the
parameter $H$ must be approximately constant. It implies that in this
regime the energy density does not change with the expansion but remains
constant too. It is indeed realized in models where inflation is driven by
a scalar (inflaton) field. Let us assume now that baryons are conserved. In
accordance with eq.(\ref{nbngamma}) the energy density associated with
baryonic charge at the hot early stage of the universe evolution
is about $10^{-10}-10^{-9}$ of the total energy density. Let us go
backward in time to even earlier period, when inflation took place. At this
stage the energy density of all forms of matter is represented by the
inflaton and remains constant in the course of contraction (remember we are
going backward in time). However the energy density associated with baryonic
charge cannot be constant because by assumption this charge is conserved.
Correspondingly it changes with the scale factor
as $\rho_B \sim a^{-4}$. It means that in less than
6 Hubble times the energy density of baryons becomes dominant and the total
energy density could not remain constant. Thus with conserved
baryons inflation can be only very short, $H_I \tau \leq 6$, which is by far
below the necessary duration.

Thus we must conclude that baryonic charge in our universe is not conserved
and the direct experimental search of the proton instability or
neutron-antineutron oscillations is not only just experiments for putting
an upper bound but the experiments for discovery really existing phenomenon.
Unfortunately cosmology does not say anything about the magnitude of the
effect.
It very much depends upon the mechanism of baryonic charge nonconservation and
one should keep in mind that the mechanism
through which the observed baryon asymmetry of the universe has been created
is not necessarily the same that leads to the proton decay or
neutron-antineutron oscillations. Theory opens several possibilities
to break B-conservation with different levels of creditability.
The standard $SU(2)\times U(1)$-electroweak interactions are known
to break baryonic current conservation by the
chiral quantum anomaly\cite{thooft}. This is a rather strong theoretical
prediction but unfortunately manifestations of this phenomenon in low energy
physics are extremely weak, they are suppressed by the tunnel penetration
factor
$\exp(4\pi\sin^2 \theta_W /\alpha) \approx  10^{-170}$. At high temperatures
the
effect may be grossly amplified and may explain the observed baryon asymmetry
of the universe\cite{krs} (for the reviews see the talk by A. Cohen at this
Conference or review papers\cite{ad1,ckn}). Fortunately there are plenty
of other mechanisms of B-nonconservation,
which do not necessarily  operate at ultrahigh
energies, as for example the GUT's one does.
Some of them are so efficient at low energies that
the direct observation of the effect is almost at hand and, as M. Goldhaber
said at the beginning of this meeting, one should rush to the laboratory and to
make the discovery (unfortunately he referred to the unsuccessful attempts to
find proton decay in the first generation experiments). Let us hope that the
second generation will make it.

\section{General conditions for cosmological creation of antimatter.}

Why at all may we expect that there are macroscopically large domains of
antimatter in the universe? There is no rigorous theory which requests that.
Moreover in all simple models of baryogenesis the baryon asymmetry is a
universal constant over all the universe so that there is no place for
antimatter. On the other hand simple modifications of baryogenesis scenarios
will result in formation of domains with different signs of baryon asymmetry.
To this end the following two conditions should be satisfied:
\begin{enumerate}
\item{}Different signs of C and CP-violation in different space points.
\item{}Inflationary (but moderate) blow-up of regions with different signs of
charge symmetry breaking.
\end{enumerate}
The first condition is realized in the model of spontaneous breaking of
charge symmetry\cite{lee}. It is assumed that the Lagrangian is charge
symmetric
but the ground state is not. It can be realized by a complex scalar field
which acquires a nonzero vacuum expectation value like
the one in the usual Higgs
mechanism. The effective potential of this field may e.g. have the form:
\beq{
U(\phi) = -m^2 |\phi|^2 +\lambda (\phi^4 + \phi^{*4}) + g^2T^2 |\phi|^2
\label{uphi}
}\eeq
where the last term came from the temperature corrections, which force the
system to the symmetric state at high temperatures\cite{kirz}. At low
temperatures the state $\langle \phi \rangle = 0$ becomes energetically
unfavorable and a complex condensate is developed which through Yukawa
coupling would give rise to
breaking of C and CP  by e.g. complex fermion masses. One
can see that through this mechanism domains with opposite signs of C(CP)-odd
phase are indeed formed. In these domains either matter or antimatter is
generated by baryogenesis\cite{brst}.
The universe in this model is charge symmetric on the
average and asymmetric locally.

There are two serious problems which this model encounters. First is that the
average size of the domains is too small. If they are formed in the second
order
phase transition, their size at the moment of formation is determined by the
so called Ginzburg temperature and is approximately equal to
$l_i=1/(\lambda T_c)$ where $T_c$ is the critical
temperature at which the phase
transition takes place and $\lambda$ is the selfinteraction coupling constant.
In this case different domains would expand together with the universe and now
their size would reach $l_0 = l_i (T_c/T_0) = 1/(\lambda T_0)$ where
$T_0 = 2.7$K is the present day temperature of the background radiation. If the
phase transition is first order then the bubbles of the broken phase are
formed in the symmetric background. In this case different bubbles initially
are not in contact with each other, typically the distance between them is much
larger than their size, and their walls may expand faster than
the universe, even as fast as the speed of light. Thus to the moment when the
phase transition is completed the typical size of the bubbles may be as large
as the horizon, $l_f\approx t \approx m_{Pl}/T_f^2 $. After that they are
stretched out by the factor $T_f/T_0$ due to the universe expansion.
To make the present day size around (or larger than) 10 Mpc we need
$T_f \sim 100 $ eV. It is
difficult (if possible) to arrange that without distorting successful
results of the standard cosmology. Thus to make observationally acceptable size
of the matter-antimatter domains, a superluminous cosmological expansion seems
necessary. This solution was proposed in ref.\cite{sato} where exponential
(inflationary) expansion was assumed. With this expansion law it is quite easy
to overfulfill the plan and to inflate the domains above the present day
horizon. Effectively it would mean a return to the old charge asymmetric
universe without any visible antimatter. So some fine-tuning is necessary which
would permit to make the domain size above 10 Mpc and below 10 Gpc.

The second cosmological problem which may arise in this model is a very high
energy density and/or large
inhomogeneity created by the domain walls\cite{zko}. This
can be resolved if domain walls were destroyed at later stage by the symmetry
restoration at low temperature or by some other mechanism \cite{ms,kts}.
However there could be scenarios of baryogenesis in which domains of
matter-antimatter may be created without domain walls. The basic idea of
these scenarios is that baryogenesis proceeds when the (scalar) field which
creates C(CP)-breaking or stores baryonic charge is not in the dynamically
equilibrium state. These models are described
in more detail in the following sections.

\section{Antimatter in models with baryonic charge condensate.}

In supersymmetric theories there exist scalar fields with nonzero baryonic
charge, superpartners of quarks. Such fields (more exactly the electrically
neutral colorless combination of squarks and sleptons)
may form a classical condensate in the early
universe, in particular at inflationary stage, if there are the so called
flat directions in the potentials. Subsequent decay of this
condensate would result in a considerable baryon asymmetry\cite{afdi}. The
picture can be visualized as follows. Evolution of a complex
spatially homogeneous scalar field is described by the same equation as
two-dimensional motion of a point-like body in the same potential
$U(Re \phi, Im \phi ) \rightarrow U(x,y)$. Baryonic charge density is
equivalent in this language to the angular momentum of the mechanical motion
of the body. The potential typically
has the form of eq.(\ref{uphi}). It is spherically symmetric at small $\phi$
and asymmetric and has flat directions at large $\phi$. So for small values of
the amplitude of $\phi$ baryonic charge is conserved while evolution of $\phi$
with a large amplitude goes with a strong baryonic charge nonconservation. If
the  mass of $\phi$ is smaller than the Hubble parameter during inflation, the
field would climb up the potential slope due to infrared instability of
scalar fields in De Sitter space-time\cite{buda,adl2,vifo}.  When inflation
ends the
field $\phi$ would evolve down to the equilibrium value. Depending upon the
initial conditions it may rotate clock-wise or anticlock-wise near the
origin or in other words it would produce baryons or antibaryons in its
decay. One sees at this example that even in the charge symmetric theory
baryon asymmetry may evolve; charge asymmetry is created by asymmetric
initial conditions which in turn are created by rising quantum fluctuations
of the scalar baryonic field during inflationary stage. Of course at large
scales the universe is charge symmetric. It is evident that there is no domain
wall problem in this scenario. The characteristic size of domain with a
definite
sign of baryonic  charge was estimated in ref.\cite{ad1}. At the end of
inflation it is equal to $L_{Bi} = H_I^{-1} \exp(\lambda^{-1/2})$. With
$\lambda$ around $10^{-3}-10^{-4}$ such domains would be consistent with
observations and still inside the present day horizon. Since it is natural
to assume that the baryon asymmetry in this model gradually changes from a
positive value through zero to a negative one, the annihilation at the
boundaries of the domains would be much weaker than in the (usually assumed)
picture of interactions of domains
with sharp boundaries. Correspondingly the limits on
the magnitude of $l_B$ would be considerably weaker. Note that not only the
sign but also the magnitude of the baryon asymmetry in different domains in
this scenario may be significantly different.

\section{Alternating (and periodic?) matter-antimatter cosmic layers.}

A relatively simple modifications of the baryogenesis scenario would permit to
get a very interesting distribution of matter and antimatter in the universe
ranging from strictly periodic flat alternating layers
of matter and antimatter\cite{doka,dikn,doch,chki}
to cell structures with each cell formed by matter or antimatter with an
average characteristic size which could easily be around 100 Mpc. The basic
assumptions leading to this kind of structure are quite simple and even
natural.
Assume that there exists a complex scalar field $\phi$ with the mass which
is smaller than the Hubble parameter at inflation, $m_\phi <H_I$. Assume also
that the potential $U(\phi)$ contains nonharmonic terms (i.e. not only
$m^2|\phi|^2$ but also e.g. $\lambda|\phi|^4$). Assume at last that a
condensate
$\langle \phi \rangle =\sigma (\vec r )$ was formed during inflation. It is
essential that the condensate $\sigma$ is not a constant but a slowly varying
function of $\vec r$. Such a condensate could be formed due to infrared
instability of the scalar field mentioned in the previous section or in first
order phase transition with very much inflated bubble walls. The characteristic
scale at which $\phi$ essentially varies, $l_\phi$, may be exponentially large
due to inflation.

When inflation is
over, the field $\phi$ relaxes down to its equilibrium value, oscillating near
the minimum of the potential. If baryogenesis takes place
very soon after the end
of the inflation and the rate of the baryogenesis is large in comparison with
the frequency of oscillations of $\phi$, then the instant value of the
amplitude of $\phi$ would be imprinted on the magnitude of the asymmetry
because, as we mentioned above, a condensate of a complex scalar field
gives rise to C(CP)-violation proportional to the field amplitude. Thus
baryogenesis makes a snapshot of the magnitude of $\phi$. Now since the
potential $U(\phi)$ is not harmonic, the frequency of the oscillations of
$\phi$
depends on the amplitude. By assumption the initial amplitude is not the
same at different space points and so the frequency is also a function of
$\vec r$. Because of that the initially smooth function $\phi(\vec r)$ would
turn into an oscillating one with a huge wave length of oscillations
proportional to $l_\phi$.

If $\phi$ oscillates around zero than its snapshot would show both positive
and negative values. In the case that there are no other comparable sources of
C(CP)-violation this model would produce approximately equal number of
baryons and antibaryons situated on relatively thin layers or shells. If the
equilibrium value of $\phi$ is nonzero or there is an explicit charge
symmetry breaking, matter or antimatter would be produced more efficiently and
the universe on the average would be more baryonic or antibaryonic.

\section{Island universe model.}

It is relatively simple to construct a cosmological model of the universe
consisting of separate baryonic or antibaryonic islands floating in
the sea of invisible matter or even of a baryonic island surrounded by the
sea of antimatter\cite{doka,dikn}. In the first case our chances to observe
antimatter are minor because the distance between the islands
is typically rather large and the probability of the collisions is low. In the
second case antimatter may be possibly observed by the gamma ray background.

In short the scenario leading  to  the  insular  structure  can  be
realized  as  follows.
First, the charge symmetry should  be  spontaneously  broken
and the phase transition to the CP-odd phase  should  be
first order with supercooling and formation of bubbles  of
the new phase inside the quasistable CP-symmetric  phase.  Second,
there should be sufficiently long period of
exponential expansion after the phase transition but not too long.
Otherwise the sizes of the  CP-odd  bubbles would be either
too small in contradiction with observations
or too large so that we would never see the boundary.
If the phase transition took place before the end of
inflation but not far from it, the island  size  could  be  of  the
order of the present horizon size but still slightly smaller  than
the latter. When inflation ends and the Universe is (re)heated  an
excess of particles over antiparticles or vice versa is  generated
inside of the bubble by the normal process of baryogenesis.
Outside of the bubbles where the charge  symmetry  is  unbroken
the baryonic charge density would be equal to zero. However it might be
that there are two mechanisms of C(CP)-breaking, the spontaneous one
operating inside the island and an explicit one operating everywhere.
In that case the baryogenesis would proceed also outside the bubbles
and may have either sign, in particular it is possible that the baryonic island
would be in the antibaryonic sea. In that case one may expect a noticeable
annihilation on the coast.

The size of the islands (or bubbles) depends upon the duration of inflation
after the phase transition to C(CP)-odd phase took place. Normally the
duration of inflation is very large in comparison with the minimal necessary
one, $H_I\tau \approx 60$, and one would naturally expect that the size would
be much larger than the present day horizon. To escape this conclusion one
may introduce a coupling of the field $\phi$, which creates charge symmetry
breaking, to the inflaton field $\Phi$, e.g. of the form:
\beq{
L_{int} =\lambda ' \mid \phi \mid ^2 (\Phi -\Phi  _1  )^2
\label{lphiPhi}
}\eeq
with $\lambda ' > 0$ and $\Phi_1$ is such that the
inflaton field reaches and passes this value in the course of inflation.
This interaction leads to effective time dependent mass of $\phi$,
$\Delta m^2 (t)=\lambda' [\Phi (t) -\Phi_1 ]^2$, so that
the state $\phi =0$ is almost always classically stable  with  respect
to small fluctuations and only when $\Phi$ is  close  to  $\Phi_1$
there is a period of instability. Quantum fluctuations  of  $\phi$
at that time  increases  and, if  they  exceed  a  critical  value
$\phi_c$ to the moment when the condition  of stability
becomes valid again, they do not return to the false vacuum
state but would rise up to  a nonzero complex value. Thus  the  bubbles  of
CP-odd vacuum can be formed. The average bubble size $d$  and  the
distance $l$ between them  are  very  much  model  dependent.  In
particular the value of $l$ can vary from $0$ to infinity and correspondingly
vary the odds for observing antimatter in such universe.

\section{Very inhomogeneous baryogenesis.}

The model considered in this section combines some of the ideas discussed
above but in an extremal form. Namely the mechanism of baryogenesis
was proposed\cite{dosi} which creates a huge baryon asymmetry
$N_B/N_\gamma = O(1)$ in relatively small regions with, say, stellar size  over
the normal homogeneous baryonic background with
$N_B/N_\gamma$ given by eq.(\ref{nbngamma}). The probability of production of
such high-B regions should be sufficiently small so that their number density
is
below the observational bounds. The sign of the baryon asymmetry in this
regions is with equal probability positive or negative so
we can expect both high density and small size baryonic and antibaryonic
objects.  There is no observational difference between the two if the density
is so high that those objects collapsed at some early epoch into compact
stellar remnants and black holes. This model presents a mechanism for
early black hole formation from large amplitude isothermal fluctuations
at small spatial scales. In this case, at least some dark matter in
the universe would be in the form of baryonic (and
antibaryonic) black holes.  Smaller uncollapsed bubbles of antibaryonic matter
would be observable either as point-like sources of $\gamma$--radiation or, if
they annihilated earlier, as some bright spots in the otherwise isotropic
background radiation.  If the number density of these objects were sufficiently
high,  early $p \bar p-$annihilation could result in the distortion of
the spectrum of background radiation.
Unfortunately there is too much freedom in the model to make any specific
predictions. The amount of
uncollapsed antimatter may vary from an unnoticeable amount to that in
contradiction with existing data.

The basic idea of the model is to make the conditions in which the
Affleck-Dine\cite{afdi} mechanism of baryogenesis could be operative
only in small spatial regions. In these regions the asymmetry may be huge
since this mechanism suffers from overabundant baryoproduction in contrast to
all other ones. This could be realized if the flat directions in
the potential of the scalar
baryonic field $\phi$ are separated from the
origin (where the field is normally
located) by a potential barrier. In this case the jump to the flat directions
could be achieved only through the tunnel transition which is usually strongly
suppressed. This ensures the desired suppression of the production of high
B-bubbles. Once again the jump to the flat directions should be done during
inflationary stage to make the bubbles macroscopically large at the present
time. The necessary tuning may be achieved by a coupling between $\phi$ and
the inflaton field.

Under reasonable assumptions about the production mechanism the mass
distribution of the high density baryonic or antibaryonic bubbles is given
by the expression\cite{dosi}:
\beq{
{ dn \over dM } = M^4_0 \exp [-\alpha- \beta \ln^2 (M/M_0) ]
\label{dndm}
}\eeq
The constants $\alpha$, $\beta$, and $M_0$ are determined by the parameters
of the potential of the $\phi$-field and the Hubble constant during inflation.
With the reasonable choice of the parameters it is possible to get $M_0$ in
the  interesting interval $(1-10^6)M_\odot$ where $M_\odot$ is the solar mass.

The cosmological evolution of such bubbles depends upon their size and the
magnitude of the baryon asymmetry or, to be more precise, upon the ratio of
their size, $l_B$ and the Jeans wave length, $\lambda_J$.
Bubbles of large size, $l_B >\lambda_J$, would form compact objects,
either stars or black holes, at a very early stage of the evolution of
the universe.
Stars of antimatter could emit considerable energy due to annihilation of
the accreted matter. With a sufficiently large amount of surrounding matter,
they should radiate at their Eddington limit,
\beq{
L_{Ed} =3\cdot 10^4 L_\odot \left( {M\over M_\odot}\right)
\label{led}
}\eeq
where $L_\odot \approx 4\cdot 10^{33}$erg/sec is the solar luminosity.
The life-time of such
objects is of the order of $5\cdot 10^9$years. If the accretion rate is
below the limiting one (e.g. due to the surrounding deficit of matter),
the luminosities would
be smaller and the life-times would be larger. Those objects can be observed as
$\gamma$-ray sources isotropically distributed over the sky. A very
interesting phenomena may
take place in a collision of the antistar with a normal star. One would expect
to observe together with a flux of gamma radiation rare events of antinuclei,
in particular anti-helium-4.

\section{Conclusion.}

One cannot make any strong conclusion from this very speculative talk. What
seems
quite definite is that baryonic charge is not conserved. Hence proton is in
principle unstable, neutron-antineutron oscillations should exist and this
is matter of "only" good luck to observe them in direct laboratory experiments.
Unfortunately cosmology is absolutely helpless in predicting the magnitude of
the  effects.

The probability of observing big lumps of antimatter in the universe suffers
from the similar uncertainty. The difference is however that in this case
the existence of antimatter is by no means obligatory. While baryonic charge
is definitely nonconserved, the universe may still contain only baryons.
Another sad but very probable option is that the universe may be charge
symmetric but antimatter is far beyond present day horizon. It means
effectively
that "our best of all possible worlds" does not contain antimatter.
Unfortunately the models with inflationary expansion of the matter-antimatter
bubbles would quite easily overexpand them too far. Still a rather simple
coupling of the underlying scalar fields to the inflaton may stop
inflating the bubbles at sufficiently early moment and make them comfortably
(for possible observations) nearby. If this is true, very interesting
configurations of matter-antimatter regions in the universe are possible, as
it has been discussed above. Anyhow independently of theoretical speculations
the idea of the charge symmetric universe looks so interesting and attractive
that the searches for that just cannot be unsuccessful.

\bigskip

This paper was supported in part by the Danish National Science Research
Council
through grant 11-9640-1 and in part by Danmarks Grundforksningsfond through its
support of the Theoretical Astrophysical Center.

\end{document}